\documentclass[12pt]{article}
\usepackage{graphics}
\begin{document}
\bigskip \bigskip \bigskip

\centerline{\large \bf Bound states and brane annihilation in 11 $D$ SUGRA}
\bigskip \bigskip
\centerline{\bf Aleksey Nudelman}
\medskip
\centerline{Department of Physics}
\centerline{University of California}
\centerline{Santa Barbara, CA  93106-9530 U.S.A.}
\medskip
\centerline{email: anudel@physics.ucsb.edu}
\bigskip \bigskip
\begin{abstract}
We study scattering of $D0$ branes in eleven dimensional supergravity (SUGRA) using the tree level four point amplitude.
The range of validity of   SUGRA allows us to make reliable calculations of relativistic brane scattering, inaccessible in string theory because of halo effects. We also compare the rate of annihilation of $D0-\bar{D0}$ branes to the rate of elastic  $D0-\bar{D0}$ scattering and find that the  former is always  smaller than the latter. We also find a pole in 
$\sigma_{D0-\bar{D0}}$. We exploit the analogy with the positronium to argue that the brane anti-brane pairs  form branium atoms in 3 spatial dimensions. We also  derive
a long range effective potential for interacting branes which explicitly depends on their polarizations. We compare two approaches 
to large impact parameter brane scattering: in GR, polarizations of the branes are kept constant during the interaction, while in  QFT, one sums over all possible polarizations. We show that the GR and QFT approaches give the same answer.
\end{abstract}
\medskip
\section{Introduction}
	The last two years have seen a revolutionary development of String Theory.
 Polchinski's realization  \cite{polch} that $D$ branes carry $R-R$ charge, and thus they are BPS objects protected by supersymmetry, had spawned a web of literature calculating interactions of $D$ branes in string theory.
It was initially noticed by Bachas \cite{bachas} that the superstring in an electromagnetic  background can be regarded as a dual description of moving $D$ branes. He also noticed that the leading order interaction between two $Dp$ branes, moving with relative velocity $v$, goes as $v^4$. Lifschytz had consecutively shown \cite{lif} that the effective potential of moving $Dp$ branes separated by distance $r$ is 
\begin{equation}
V(v,r) \sim -v^4/r^{7-p}
\end{equation}
He also found a potential for long distance interaction between the $Dp$ brane and the $\bar Dp$ anti-brane
\begin{equation}
V(v,r) \sim -(2+v^2)/r^{7-p}
\end{equation}
As was later argued by Banks and Susskind \cite{force} the force of attraction between the $D0$ brane and the $\bar D0$ anti-brane diverges when 
\begin{equation}
r \sim 1/\sqrt{\alpha'} 
\end{equation}
The purpose of this paper is threefold:
Firstly, we use Witten's idea  \cite{witten} that $D0$ branes are just Kaluza-Klein gravitons of the eleven dimensional supergravity to study the  brane interaction as a problem of tree level quantum gravity. We explicitly include brane polarizations in our 
calculations.

Secondly, we scatter $D0-\bar{D0}$ for different values of momenta. At small momentum we probe distances much larger than the Plank scale and cross sections
vanish. As we increase the momentum the brane annihilation catches up with elastic 
 scattering and we expect it to become dominant near the Planck scale  where SUGRA description breaks down.
Thirdly we study long range  brane scattering and find that the functional form of the amplitude remains the same for $D0-D0$ brane scattering for small and for large incoming transverse momenta. As expected, the $D0-\bar{D0}$ interaction levels off with the  $D0-D0$ interaction for large momenta.

\section{Historical digression on Quantum Gravity.}
	Our task is greatly simplified by  Sannan's \cite{sannan} calculation of four-point amplitude in quantum gravity which is independent of the number of dimensions.
Since the success of any given quantum gravity calculation is limited, it is hard to trace 
earlier attempts to quantize gravity. The father of "modern" quantum gravity, Richard Feynman, \cite{feynman} refused to publish his results for 10 years. It had been 7 years and 30 citations after Bryce DeWitt \cite{dewitt} published three and four graviton vertex functions when Berends and Gastmans \cite{berends} found an error in his 360 term four graviton vertex function. Berends and Gastmans, as well as Feynman, used computers to make algebraic calculations. It is curious to note that Sigurd Sannan, 20 
years and 200 citations later, found one more mistake in  DeWitt's original calculation.
Sannan's graviton - graviton scattering amplitude (calculated by hand) contains 150 terms (it would take four pages size of Physical Review to write it down) 

	 Before we undertake the study of graviton scattering, we should try to understand how it may depend on the polarization. The graviton  polarization itself
is not  very well defined in the literature. Killing symmetry becomes a gauge symmetry for linearized GR. Graviton self interaction terms are not gauge invariant. Also the whole concept of perturbation theory is not  defined since the Minkowski metric around which we perturb is a pure gauge from the point of view of gauge theory. However a good  argument can be made from the GR point of view. Let us consider an approximate solution to  the Einstein's equation 
\begin{equation}
g_{\mu \nu}=\eta_{\mu\nu}+e_{\mu \nu} e^{ikx}
\end{equation}
 where 
\begin{equation}
e_{\mu \nu}e^{\mu \nu}<<1
\label{privet}
\end{equation}
The energy density of this configuration is 
$T_{00} \sim E^2e^2/l_p^8$.
Now let us impose a self-consistency condition on (\ref{privet}): we want this energy density when put on the right hand side  of the Einstein's equation to create gravitons with polarization of the same magnitude  as the original one. The ``mean field''
equation is
\begin{equation}
e \sim e^2 /(Er)^2
\end{equation}
From which it follows that $rE>>1$. The limit of linearized GR where any graviton interactions, including that of the spin flip , are suppressed is just Born's approximation.  
\section{Long Range Brane Interactions.}
	In this section  we find physical  limits of Sannan's $150$ term amplitude which reduce it to $6$ terms. A way to proceed is to remember that the completeness
relation for polarizations of the photon in QED gets drastically simplified once we choose its space momentum along one of the coordinate axes.
Since the completeness relation is proportional to a photon propagator, one finds  great simplifications in the photon scattering amplitudes as well.
A natural way to single out a space direction is to consider scattering of branes with small transverse momentum $q$. In this limit the natural dimensionless parameter is $qR_{10}$. For small $qR_{10}$, we are dealing with  forward scattering in the compact
 direction. For small $\frac{1}{qR_{10}}$ it   is a forward scattering in a non-compact direction. Since in both limits the momentum exchange is infinitesimal, it is not surprising that we get the same functional form for the amplitude.
The details of the construction are as follows:

The graviton has $d(d-3)/2$ different polarizations in $d$ dimensions. In $d$ dimensions it is exactly the number of independent components in the traceless symmetric  matrix $d-2$ by $d-2$.
For a $D0$ or $\bar D0$ brane "moving" only in the compact 10-th direction, the 11-momentum  is
\begin{equation}
k^m=(E,0,0,..0,\pm 1/R_{10})
\label{mom}
\end{equation}
We can choose the polarization as the following matrix
\begin{equation}
\left (\begin{array}{ccccc}
0 & 0& \ldots  &0& 0 \\
0 &  & &  & 0 \\
\vdots & & & & \vdots \\
0 &  & &  & 0 \\
0 & 0& \ldots  &0& 0
\end{array} \right)
\label{pol}
\end{equation}
, where all the empty spaces inside of (\ref{pol}) correspond to some arbitrary numbers.
At this stage, if we have four different gravitons, they may all have different polarizations, corresponding to different matrix elements of (\ref{pol}).  However, the 11-momentum (\ref{mom}) of i-th graviton is transverse to the polarization of the j-th graviton.
\begin{equation}
k^i_m e^j_{mn}=0
\label{trans}
\end{equation} 
	This observation is technically very important since it reduces Sannan's amplitude (formula (3.16) of \cite{sannan})
to only six terms.
Now we can make a rotation in the $x_{10} x_{9}$ plane and allow "a little bit of momentum" in the non-compact directions. The matrix (\ref{pol}) will receive corrections of order $\sim R_{10}q$ where q is the magnitude of non-compact momentum. As long as
  $qR_{10}$ is small the leading order terms of the amplitude expansion in terms of $qR_{10}$  will contain only six terms as mentioned above. In other words, for small $qR_{10}$,  we can approximately take the 11-momentum  of $i$-th graviton to be
 transverse to the polarization of the $j$-th graviton. This corresponds to non-relativistic brane scattering.
Just before we list the advertised 6-term amplitude, we should mention yet another possibility.
Let us start with four gravitons which have momentum in some non-compact direction and none in the compact. These gravitons will  have polarization matrices of the form (\ref{pol}), satisfying mutual transversality  condition (\ref{trans}). Now, let us make a rotation to allow a little bit of compact momentum $1/R_{10}$. It corresponds to having small corrections to the amplitude of order $1/(qR_{10})$. Our branes are now  fully relativistic.\footnote{We discuss validity of Born's approximation for relativistic branes at the end of the chapter 4}
The convenient choice of polarization has allowed us to get the following amplitude for small and large values of non-compact momenta:
\begin{eqnarray}
T=-1/4 \kappa^2 \left(\frac{st}{u} (e_1e_3)(e_2e_4)+\frac{su}{t}(e_1e_4)(e_2e_3)+\frac{tu}{s}(e_1e_2)(e_3e_4)\right)- \nonumber \\
1/2\kappa^2 \left(s(e_1e_4e_2e_3)+t(e_1e_2e_4e_3)+u(e_1e_2e_3e_4) \right)
\label{amp}
\end{eqnarray}
We use Sannan's notation with 
\begin{equation}
S=\frac{2}{\kappa^2} \int d^{11}x \sqrt{-g} R
\end{equation}
The reason why for obtaining the same functional form of the amplitude for ultra relativistic and non-relativistic is a simple one: Choice of polarization (\ref{pol}) allows us to treat 
gravitons as particles of mass $1/R_{10}$ in one limit and of mass $q$ in the other.
 
For the large impact parameter scattering, the $t$ channel  dominates and in the limit  $t \rightarrow 0$, we get the  the following  amplitude
\begin{equation}
T_{long range}=-1/4 \kappa^2 \frac{su}{t} (e_1e_4)(e_2e_3)
\end{equation}
For $D0-D0$ brane scattering we take graviton 11-momenta as
\begin{eqnarray}
k^m_1&=&(E,0,0,...,0,q,1/R_{10}) \nonumber \\
k^m_2&=&(E,0,0,...,0,-q,1/R_{10}) \nonumber \\
k^m_3&=&(-E,0,0,...,q\sin(\theta),q\cos(\theta),-1/R_{10}) \nonumber \\
k^m_4&=&(-E,0,0,...,-q\sin(\theta),-q\cos(\theta),-1/R_{10}) 
\label{mog}
\end{eqnarray}
where $E=\sqrt{q^2+1/R_{10}^2}$. The $D0-\bar D0$ momenta is obtained from (\ref{mog}) by switching signs of compact momenta of the second and third gravitons. 
For large impact parameter, the $\theta$ is small and the amplitudes become
\begin{equation}
T_{D0-D0}=-\frac{4 \kappa^2 q^4}{q^2 \theta^2}(e_1 e_4) (e_2 e_3)
\label{amp1}
\end{equation}
\begin{equation}
T_{D0-\bar D0}=-\frac{4 \kappa^2 E^4}{q^2 \theta^2}(e_1 e_4) (e_2 e_3)
\label{amp2}
\end{equation}
For small momenta $q$ the $D0-\bar D0$ scattering dominates that of $D0-D0$ and for large momenta they are equal. This is what we could have  expected since fast 
moving branes which have an impact parameter so large that they almost do not deflect are charge blind. 
As discussed in chapters 2 and 4,  polarizations of a brane should be set equal before and after the scattering.   
\section{Low energy effective potential} 
Let us normalize the Green's function so that it satisfies the equation:
\begin{equation}
\Delta^2 G(\vec{x}-\vec{y})+k^2 G(\vec{x}-\vec{y})=-\delta (\vec{x}-\vec{y})
\end{equation}
In $9$ spatial dimensions, the retarded Green's function is
\begin{equation}
G(qr)=\frac{e^{iqr} w(qr)}{r^7} 
\end{equation}
where
$w(qr)=\frac{i}{32 \pi^4} (-15i-15qr+6i(qr)^2 +(qr)^3)$. 
The Born's formula becomes
\begin{equation}
f(q,\theta)=-\frac{i m q^3}{16 \pi^4} \int{d^9 \vec{x}U(\vec{x})e^{-i\vec{k} \vec{r}}}
\label{born}
\end{equation}
where $k=2 q \sin(\theta/2)$, $m$ is the reduced mass of two zero branes and $f(q,\theta)$, the quantum mechanical amplitude, is defined in terms of differential cross-section as
\begin{equation}
 |f(q,\theta)|^2=\frac{d\sigma}{d\Omega}
\label{f}
\end{equation}
On the other hand, the center of mass differential cross-section in $9+1$ QFT is
\begin{equation}
d\sigma=\frac{|T'|^2 p_f q^6 d\Omega}{(2 \pi)^8 p_i 64 E^2}
\label{qft}
\end{equation}
 where $T'$ is $11$ dimensional scattering amplitude compactified to $10$ dimensions, and 
$p_i$ and $p_f$ are the initial and final non-compact momenta which are equal, except for processes of brane creation or annihilation. 
The compactification to 10 dimensions is done by replacing $11$ dimensional Newton's constant $\kappa^2$ by ten dimensional  $\kappa_{10}^2$, where $\kappa_{10}^2=\frac{\kappa^2}{2 \pi R_{10}}$, which relates the amplitudes:

\begin{equation}
T'=\frac{T}{2\pi R_{10}}
\label{com}
\end{equation}
Comparing formulas (\ref{f}) and (\ref{qft}), and using (\ref{com}), one can read off the relation between $T$ and $f$ up to a phase factor.
This gives us 
\begin{equation}
U(\vec{x})=-\frac{1}{16 \pi m E R_{10}} \int {d^9 \vec{k}T(\vec{k})e^{i\vec{k} \vec{x}}}
\label{potential}
\end{equation}
For large impact parameter scattering, formula (\ref{potential}) gives us:
\begin{equation}
U_{D0-D0}=-\frac{15 \kappa^2 q^4 (e_1 e_4) (e_2 e_3)}{2 (2\pi)^5 E r^7}
\end{equation}
, which reduces for small $q$ to the result previously derived by \cite{fuckers,bekers} if the polarization
of the $D0$ branes does not change during the interaction.(In  reference \cite{fuckers}, the  formula for the effective potential is given in terms of string tension and  brane ``velocity''. To get the same numerical factor one should note that our definition
 of Newton's constant differ by a factor of 4 \footnote{There seems to be an equal number of authors using either type of notation for $\kappa$. We believe that 
Sannan's notation, which we use in this paper, is more natural since the kinetic part of the Lagrangian has a conventional factor of $1/2$.} and that we consider gravitons with only one unit of compact
 momentum. Using  momenta given by the formula (\ref{mog}) we get the same effective potential in the limit of compact forward scattering, including the numerical factor). Alternatively, one can think that polarizations of branes might change during the 
interaction  but our picture of their interaction is a bit fuzzy which corresponds to averaging over all  polarizations. In this slightly  rotated reference frame,
the completeness relation is very simple:
\begin{equation}
\sum_{p=1}^{44} e^p_{ij}e^p_{kl}=\frac{1}{88}(\delta_{ik}\delta_{jl}+\delta_{il}\delta_{jk}-\frac{2}{9}\delta_{ij}\delta_{kl})
\label{compl}
\end{equation}
, where indices $i,j=1,2,...9$ run over the "inside" of the matrix (\ref{pol}).
The averaging is done with the aid of the following formula
\begin{equation}
\left <(e_1 e_4) (e_2 e_3) \right>=\sqrt{(e_1 e_4)^2 (e_2 e_3)^2}/2
\label{az}
\end{equation}
where factor of two appears to compensate for over counting, since nothing is changed if we  interchange polarizations of first and fourth branes with that of the second and third. With the use of (\ref{compl}) we find that the right hand side of (\ref{az}) is $1$. Therefore averaging over polarizations gives the same result as  keeping the polarization of a given brane the same before and after the interaction.
The corresponding potential for $D0-\bar{D0}$ interaction is
\begin{equation}
U_{D0-\bar{D0}}=-\frac{15 \kappa^2 E^3 (e_1 e_4) (e_2 e_3)}{2 (2\pi)^5 r^7}
\end{equation}
Let us note that one can easily derive higher order corrections to the  effective potential from the all-channel amplitude which would 
give, in general, higher powers of $1/r$ with appropriate powers of momenta. Let us stress again that loop corrections to the effective potential are suppressed as long as $r$ is larger than the Planck scale. To be more precise let us consider the  quantum gravity action
\begin{equation}
S \sim \int{d^{11}x \sqrt{-g}(R+l_p^2 R^2+\dots)}
\end{equation}
For long distance interactions, the  $t$ channel dominates
and the amplitude from the first term goes  as 
$T\sim \frac{1}{k^2}$,
where  $k=2q\sin{\theta/2}$ is (a very small)  exchanged momentum.
Since loop corrections to the GR actions go roughly as $\int{\frac{dk^{10}}{k^2+t}}$ the  quantum correction to the amplitude is  
$l_p^2 +k^2 +O(k^4)\dots$
Therefore in the limit of forward scattering $Er>>1$ , quantum gravity corrections are totally negligible.

\section{Branium}
In this section, we discuss a  possibility of a brane anti- brane atom which we 
call Branium.
Bound states in scattering problems appear as the angle independent poles of 
the scattering amplitude. We therefore limit our considerations to the poles
of the $s$ channel. The most general incoming momenta are
\begin{eqnarray}
k^m_1&=&(E_1,0,0,...,0,q,N_1/R_{10}) \nonumber \\
k^m_2&=&(E_2,0,0,...,0,-q,N_2/R_{10}) 
\end{eqnarray}
The condition for $s$ to vanish reads 
\begin{eqnarray}
N_1=-N_2 \nonumber \\
q=\frac{\mp iN_1}{R_{10}}
\end{eqnarray}
The situation is completely analogous to the  electron-electron and electron-positron scattering as described by \cite {blp}. The electron- electron amplitude has no angle independent poles and electron-positron amplitude has a pole at threshold where $E^2=0$
. This pole however does not correspond, in the case of electron-positron scattering, to a positronium bound state and just indicates possible electron-positron annihilation. Instead, the existence of the positronium is
derived from the long range Coulomb potential between an electron and positron.
In the region where the bound state is formed higher order corrections are small and correspond to fine structure shift of the spectrum. 
$11$ dimensional supergravity is  valid at distances larger than a Planck Scale. The  long range effective potential between a  brane and anti-brane is Coulomb. However the Coulomb potential  is not binding in $9$ spatial dimensions. If the positronium analogy is correct
, and strong similarities of scattering amplitudes as well as cross sections (see derivations in section 6) suggests that it is, then we should not expect any new physics to appear at  the intermediate scale (still much larger than the Plank scale where the annihilation is suppressed but much shorter than the Coulomb region, so that corrections to the Coulomb interaction are significant) and the Branium atom does not exist. One can turn this argument around to see that compactifing to  $3$ spatial dimensions (and neglecting all the singularities that might appear) 
one finds a binding Coulomb potential and thus Branium exists there, with the first Bohr's orbit of radius $\frac{R_{10}^{19}}{l_p^{18}}$. Let us note that at these distances  
quantum gravity corrections, and in particular the ladder four point diagrams  can be neglected  and tree level results are a good framework to study the  possibility of bound states. 
\section{Pair creation}
As yet another exercise in supergravity, let  us  scatter  two Kaluza-Klein
particles to create ordinary gravitons.(Creation cross section is of course the same as the annihilation cross section up to a phase factor $q^2/E^2$)
\begin{eqnarray}
k^m_1&=&(E,0,0,...,0,E,0) \nonumber \\
k^m_2&=&(E,0,0,...,0,-E,0) \nonumber \\
k^m_3&=&(-E,0,0,...,q \sin(\theta),q \cos(\theta),-N/R_{10}) \nonumber \\
k^m_4&=&(-E,0,0,...,-q \sin(\theta),-q \cos(\theta),N/R_{10})
\end{eqnarray}
To simplify calculations, we will choose the same  polarization (the inside of (\ref{pol}) is taken as an arbitrary $7$ by $7$ matrix) for all gravitons. Then, the mutual transversality condition (\ref{trans}) can be imposed, and
Sannan's amplitude then takes the form 
\begin{equation}
T=-1/4 \kappa^2 \left(\frac{st}{u}+\frac{su}{t}+\frac{tu}{s}\right)
\end{equation}
The differential cross-section, as defined by eqs (\ref{qft}-\ref{com}) is
\begin{equation}
d\sigma_{annihilation}=\frac{\kappa^4 }{3 R_{10}^2 (qE)(4 \pi)^6} \frac{q^6 (3E^2+ q^2 \cos(\theta)^2)^4 \sin(\theta)^7 d\theta}{16^2 (E^2-q^2 \cos(\theta)^2)^2}
\end{equation}
Working in higher spatial dimensions gives us a natural regularization of the amplitudes and renders full cross-sections finite:
\begin{eqnarray}
\label{anih}
\sigma_{creation}&=&\frac{q^2 \sigma_{anihilation}}{E^2} \\
\sigma_{annihilation}&=&\frac{\kappa^4}{3 R_{10}^2 (qE)(4 \pi)^6} (I_1 +\frac{m^4}{\sqrt{q^2+m^2}} I_2 \arctan[\frac{q}{\sqrt{q^2+m^2}}]) \nonumber \\
I_1&=&-7 m^{10}-\frac{67 m^8 q^2}{3} -\frac{371 m^6 q^4}{15}-\frac{555 m^4 q^6}{56}+\frac{16 m^2 q^8}{315}+\frac{278 q^{10}}{495}\nonumber \\
I_2&=&\frac{7 m^{8}}{q}+27 m^6 q+25 m^2 q^5 +6q^7+39m^4q^3 \nonumber \\
m=\frac{1}{R_{10}} \nonumber
\end{eqnarray}
For a generic scattering process, all the terms in the cross section give contributions of the same order of magnitude. These cross sections can be trusted as long as the energies and momenta involved are smaller than Plank mass.
Near threshold, $q \rightarrow 0$, the cross section to annihilate  $D0-\bar{D0}$  pair goes to zero as $q^5$. In fact the expression inside the brackets of (\ref{anih}) for small $q$ is $\frac{81 m^4 q^6}{280}$. For the annihilation cross section we get 
\begin{equation}
\sigma_{annihilation}=\frac{27 \kappa^4 m^5 q^5}{(4\pi)^6 280}
\label{net}
\end{equation}

Eleven dimensional Plank scale is conventionally normalized \cite{sannan}
\begin{equation}
\kappa^2=\frac{2\pi g^2}{m} 
\end{equation}
where $g$ is the string coupling constant.
In string units, we have
\begin{equation}
\sigma_{annihilation}=\frac{27 q^5 g^4 m^3}{1120 (4\pi)^4}
\end{equation}
The opposite process, creation of branes from gravitons,  goes as $q^7$. In both cases  the cross section is a monotonously growing function of $q$.
For comparison, we calculate full cross-sections for the processes considered  in section 4.
\begin{equation}
\sigma_{D0-D0}=\frac{278}{1485} \frac{\kappa^4 q^{10}}{(4 \pi)^6(E R_{10})^2}
\end{equation}
\begin{eqnarray}
\label{cole}
\sigma_{D0-\bar{D0}}&=&\frac{\kappa^4}{3(E R_{10})^2 (4\pi)^6} (A+B+C)\\
A&=&8 E^{10}-\frac{64 E^{16}}{q^6}+\frac{136 E^{14}}{q^4}-\frac{244 E^{12}}{3 q^2}+\frac{8 E^8 q^2}{5}+\frac{12 E^4 q^6}{35}-\frac{4E^2q^8}{35} \nonumber \\
B&=&\frac{5 q^{10}}{63}-\frac{2 q^{12}}{105 E^2}+\frac{q^{14}}{165 E^4}-\frac{64 E^{18} \log[1-\frac{q^2}{E^2}]}{q^8}+\frac{168 E^{16} \log[1-\frac{q^2}{E^2}]}{q^6} \nonumber \\
C&=&-\frac{144 E^{14}\log[1-\frac{q^2}{E^2}]}{q^4}+\frac{40 E^{12}\log[1-\frac{q^2}{E^2}]}{q^2}\nonumber
\end{eqnarray}
where $\sigma_{D0-\bar{D0}}$ stands for elastic brane-anti-brane scattering.
For small q cross-section (\ref{cole}) goes as $q^2$
The cross-section (\ref{cole}) can be analytically continued to imaginary values of $q$. The negative cross-section is again a monotonously growing function of momenta. Let us   pause here and ask: Do we trust these cross sections
for arbitrary $q$ and $m$, or for large $q$, do quantum gravity corrections
become important? Since the cross section varies roughly as $r^8$ and  perturbation theory is valid for $rE>>1$, the above cross sections are valid if $\sigma>>1/E^8$. 

\begin{figure}
\begin{center}
\caption{$\frac{\sigma_{anihilation}}{\sigma_{D0-\bar{D0}}}$ versus  $q$ for $R_{10}=1$}
\scalebox{0.5}{\includegraphics{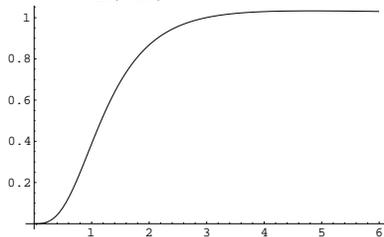}}
\end{center}
\end{figure}
\section{Conclusion}
Let us give an overview of our results.
We started the paper by studying large impact parameter brane scattering.
The amplitude for the arbitrary polarization is extremely complicated and we
limited  ourselves to almost forward scattering. We found that in the limits 
of very fast and very slow moving branes the amplitude took the same functional form. We  also explained why polarization of the branes should not change during the interaction, at least to leading order in the large impact parameter expansion. On the other hand one  can get the same answer by considering  repeated  brane scattering for different polarizations and then taking the average. We derived an effective potential for brane interactions with the aid of the retarded Green's function and matching of the QFT and Quantum Mechanical scattering amplitudes. We also discussed the possibility for the formation
of the brane -anti-brane atom and by analogy with positronium we  suggested  that
it might be  possible only in 3 spatial dimensions. We have derived cross
sections for  brane -anti-brane pair creation from two gravitons as well
as for the annihilation of the brane anti-brane pair. Consistent
with the understanding that supergravity is a long distance limit of $M$ theory,
the cross-section for brane creation is of  the  same order of magnitude as that 
for elastic brane scattering, which essentially means that we are in the regime of mostly elastic scattering.

\subsection*{Acknowledgements}

I thank Steve Giddings for numerous conversations and encouragement. I
 would also like to thank Per Kraus, Phillipe Pouliout and especially Gary Horowitz and  Joe Polchinski for discussions.
I am grateful to Smitha Vishveshwara for proofreading.

\end{document}